\documentclass[aps,prb,twocolumn,superscriptaddress,citeautoscript,footinbib]{revtex4-2}
\usepackage[dvips]{graphicx}
\usepackage{amsmath,amssymb}
\usepackage{braket}
\usepackage{float}
\usepackage{here}
\usepackage{bm}

\newcommand*{\vek}[1]{{\boldsymbol #1}}

\begin{document}
\title{Ground-state phase diagram of the one-dimensional $t$-$J_s$-$J_\tau$ model at quarter filling}

\date{\today}

\author{Yuya Kurebayashi}
\email{kurebayashi@cmpt.phys.tohoku.ac.jp}
\affiliation{Department of Physics, Tohoku University, Sendai 980-8578, Japan}
\author{Hiroki Oshiyama}
\affiliation{Graduate School of Information Science, Tohoku University, Sendai 980-8579, Japan}
\author{Naokazu Shibata}
\affiliation{Department of Physics, Tohoku University, Sendai 980-8578, Japan}

\begin{abstract}
 We study the ground state of the one-dimensional ``$t$-$J_s$-$J_\tau$ model,''
 which is a variant of the $t$-$J$ model with an additional channel degree of freedom.
 The model is not only a generalization of the $t$-$J$ model but also an effective model
 of the two-channel Kondo lattice model in the strong-coupling region.
 The low-energy excitations and correlation functions are systematically calculated by the 
 density matrix renormalization group method, and the ground-state phase diagram at quarter filling 
 consisting of a Tomonaga-Luttinger liquid, 
 spin-gap state, channel-gap state, insulator, and phase separation is determined.
 We find that weak channel fluctuations stabilize the spin-gap state, while 
 strong channel fluctuations lead to the transition to the insulator.
\end{abstract}
\maketitle

\section{INTRODUCTION}
Quantum fluctuations are important features
of microscopic systems, which give rise to plenty of interesting phenomena.
In condensed matter physics, spin fluctuations play an important role
in realizing various quantum states,
such as spin liquids and superconductivity.
One of the minimal theoretical models containing both spin and charge
degrees of freedom is the $t$-$J$ model.
The model was originally proposed
to describe high-$T_{\mbox{\scriptsize c}}$ superconductivity \cite{cite:zhang_rice},
and its one-dimensional model has been studied 
to understand the fundamental properties of strongly correlated systems.
Although this model contains only the kinetic energy term and the exchange 
energy term, various quantum states including a spin-gap state are realized \cite{cite:tj_1d_pd},
and it is interesting to investigate whether new quantum states appear
when we include additional interactions existing in more realistic systems.
One simple extension is the inclusion of repulsive interaction $V$ between 
the neighboring electrons. 
It has been reported that the repulsive interaction $V$
stabilizes the spin-gap phase at quarter filling \cite{cite:troyer}, but
other new states have not yet been obtained.
Another approach to extend the $t$-$J$ model is to add new degrees of freedom of electrons.

Praseodymium contained in cage-shaped composites, such as $\mathrm{PrTi_2Al_{20}}$,
has a non-Kramers doublet as the crystal-field ground state \cite{cite:cox,cite:onimaru}.
The theoretical model of such materials is
the two-channel Kondo lattice model (TCKLM)\cite{cite:nozieres, cite:schauerte}, which has
multiple degrees of freedom associated with the non-Kramers doublets.
As one of the simplest models of interacting electron systems consisting of multiple 
degrees of freedom, we propose the ``$t$-$J_s$-$J_\tau$ model,'' which is not only 
an extension of the $t$-$J$ model but also an effective model of the TCKLM in 
the strong-coupling region.

In this paper, we study the ground-state properties of the model by the density matrix renormalization group (DMRG) method and investigate the 
effect of the channel degree of freedom on the ground state.
The obtained results show that the spin-gap state is stabilized by weak 
channel fluctuations, while strong channel fluctuations lead to the 
transition to the insulator.

\section{MODEL}
The model we study here is the following $t$-$J_s$-$J_\tau$ model:
\begin{align}
 H_{tJJ} &= -t\sum_{i \alpha \sigma}
 \left(
 a^\dagger_{i\sigma}b_{i\alpha}b^\dagger_{i+1,\alpha}a_{i+1,\sigma}
 + \mathrm{H.c.} \right) \nonumber \\
 &+ J_s\sum_{i} \vek{S}_i \cdot \vek{S}_{i+1}
 + J_\tau\sum_{i} \vek{\tau}_i \cdot \vek{\tau}_{i+1}
 + V\sum_{i} n_in_{i+1},
 \label{eq:effective-model}
\end{align}
where $a^\dagger_{i\sigma}$ and $b^\dagger_{i\alpha}$ are the creation operators of
particles and ``holes'' at the $i$th site
with spin $\sigma$ and channel $\alpha$, respectively,
and the empty and double occupancies of $a^\dagger_{i\sigma}$ and $b^\dagger_{i\alpha}$
are inhibited:
\begin{align}
 \sum_\sigma a^\dagger_{i\sigma}a_{i\sigma} 
 + \sum_\alpha b^\dagger_{i\alpha}b_{i\alpha} = 
 n_i + \sum_\alpha b^\dagger_{i\alpha}b_{i\alpha} = 1,
 \label{eq:ed-condition}
\end{align}
where $n_i=\sum_\sigma a_{i\sigma}^\dagger a_{i\sigma}$ is
the number operator of the particles.
The spin and channel-pseudospin operators are 
defined as $\vek{S}_i = \frac{1}{2}\sum_{\sigma\sigma'}
a^\dagger_{i\sigma}\vek{\sigma}_{\sigma\sigma'}a_{i\sigma'}$ and
$\vek{\tau}_i = \frac{1}{2}\sum_{\alpha\alpha'}
b^\dagger_{i\alpha}\vek{\sigma}_{\alpha\alpha'}b_{i\alpha'}$,
respectively.

\begin{figure}[t]
 \centering
 \includegraphics[scale=1.0]{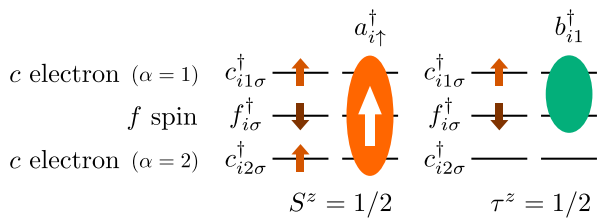}
 \caption{Schematic representations of the composite particles
 $a^\dagger_{i\uparrow}$ and $b^\dagger_{i1}$ in the TCKLM.}
 \label{fig:basis}
\end{figure}

This model is not only a generalization of the extended $t$-$J$ model 
but also derived from the TCKLM
\begin{align}
 H_{\mbox{\scriptsize TCKLM}} &= -\tilde{t}\sum_{i\sigma\alpha}\left(
 c^\dagger_{i\alpha\sigma} c_{i+1,\alpha\sigma} + \mathrm{H.c.}
 \right) \nonumber \\
 & ~ ~ + \frac{J}{2}\sum_{i\alpha\sigma\sigma'} \tilde{\vek{S}}_i\cdot
 \left(
 c^\dagger_{i\alpha\sigma}\vek{\sigma}_{\sigma\sigma'}c_{i\alpha\sigma'}
 \right), \label{eq:tcklm}\\
 \tilde{\vek{S}_i} &= \frac{1}{2}\sum_{\sigma\sigma'}f^\dagger_{i\sigma}\vek{\sigma}_{\sigma\sigma'}f_{i\sigma'},
\end{align}
 as follows:
assuming that the number of conduction electrons per local spins $n_c$
satisfies $1 \leq n_c \leq 2$,
the effective Hamiltonian in the strong-coupling region is given
by the second-order perturbation of $1/J$ from the limit $J/\tilde{t} = \infty$.
In this case, $a^\dagger$ and $b^\dagger$ are
the composite particles defined as
$ a^\dagger_{i\sigma} = \frac{1}{\sqrt{6}}
 \left(
 2c_{i1\sigma}^\dagger c_{i2\sigma}^\dagger f_{i\bar{\sigma}}^\dagger
 - c_{i1\sigma}^\dagger c_{i2\bar{\sigma}}^\dagger f_{i\sigma}^\dagger
 - c_{i1\bar{\sigma}}^\dagger c_{i2\sigma}^\dagger f_{i\sigma}^\dagger
 \right)$
and
$ b^\dagger_{i\alpha} = \frac{1}{\sqrt{2}}
 \left(
 c_{i\alpha\uparrow}^\dagger f_{i\downarrow}^\dagger
 - c_{i\alpha\downarrow}^\dagger f_{i\uparrow}^\dagger
 \right)
$
, respectively, as schematically represented in Fig. \ref{fig:basis}.
The transfer integral $t$ in Eq.~(\ref{eq:effective-model}) is given
as $t=3\tilde{t}/4$ and
the effective interactions are
\begin{align}
 J_s = \frac{1504t^2}{135J}, \ 
 J_\tau = \frac{64t^2}{9J}, \ \ \
 \frac{J_\tau}{J_s} \approx 0.64
 \label{eq:interaction}.
\end{align}
The interactions $J_s$ and $J_\tau$ are the
two largest terms obtained by the perturbation expansion, and
the other ones, including next-nearest hopping, are ignored.
The neglected long-range interactions are expected to suppress 
the phase separation caused by the above two interactions.
Instead of treating all such terms explicitly,
we consider the repulsion term $V$ to suppress the phase separation.
Note that when $n_c=1$, 
which corresponds to the absence of the $a^\dagger$ particles ($n=0$),
the model is reduced to the Heisenberg model
of the channel degree of freedom \cite{cite:schauerte}.

In this study, we analyze the ground state of the Hamiltonian of Eq.~(\ref{eq:effective-model})
with an equal number of particles and holes at 
$n=1/2$ (quarter filling, $k_F=\pi/4$). 
This filling corresponds to $n_c=3/2$ in the TCKLM.
Throughout this paper, we fix the nearest-neighbor interaction $V$ as $V/t = 0.8$
and take the transfer integral $t$ as the unit of energy.

\section{METHOD}
We use the DMRG method \cite{cite:dmrg_white1, cite:dmrg_white2}
to analyze the ground states of the Hamiltonian of Eq.~(\ref{eq:effective-model}).
In this method, the accuracy of the ground-state wave function
is systematically controlled by the number of remaining states $m$. 
We increase $m$ up to 400 to see the convergence of the 
results, where the truncation error is less than $10^{-5}$.
The system size is in the range of 128--192.

To suppresses the finite-size effect caused by
the open boundary conditions used in the DMRG calculation,  
we apply the sine square deformation (SSD)
\cite{cite:ssd_gendiar1, *cite:ssd_gendiar2} to the Hamiltonian.
Since the SSD reproduces the bulk response
to an external field \cite{cite:ssd_hotta},  
we use this property to obtain the excitation gap 
of the infinite system.

\normalsize
\begin{figure}[H]
 \centering
 \includegraphics[scale=1]{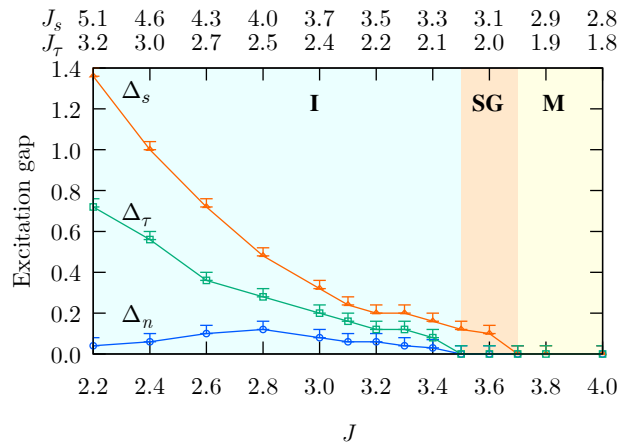}
 \caption{Excitation gaps for
 the charge $\Delta_n$, spin $\Delta_s$, and channel $\Delta_\tau$ degrees of freedom.
 The upper axes represent the effective interactions defined in Eqs.~(\ref{eq:interaction}).
 The error bars are introduced by the discrete parameter settings used in the SSD method \cite{cite:ssd_hotta}. 
 }
 \label{fig:ext_gap}
\end{figure}
\begin{figure}[H]
 \centering
 \includegraphics[scale=1]{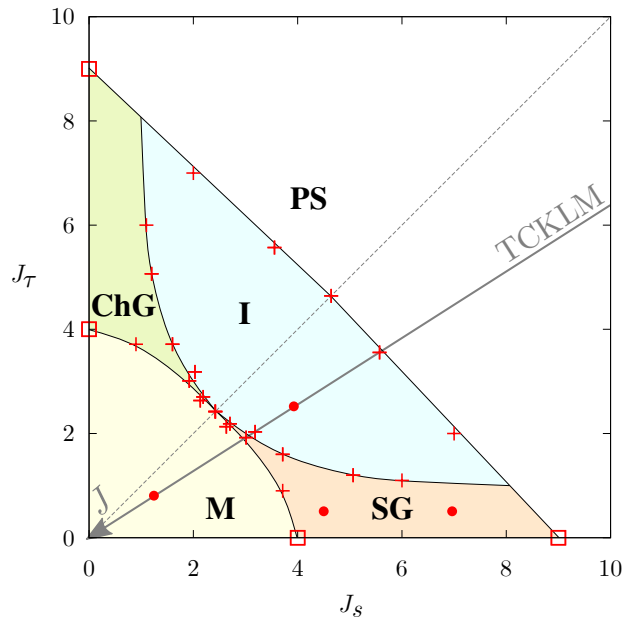}
 \caption{Ground-state phase diagram of the $t$-$J_s$-$J_\tau$ model.
 The plus marks represent the transition points.
 The phase boundary is roughly drawn. 
 The squares on the transition line were determined in a previous work \cite{cite:troyer}
 for the extended $t$-$J$ model.
 The circles in the phase diagram represent the points where
 the correlation functions shown in
 Figs.~\ref{fig:metal_long_corr}, \ref{fig:sg_long_corr}, and \ref{fig:insu_long_corr} are calculated.
 M, metallic phase (no excitation gap); SG, spin-gap phase (only the spin gap opens); ChG, channel-gap phase (only the channel gap opens); I, insulating phase (gap opens for all excitations); and PS, phase separation.}
 \label{fig:phase_diagram}
\end{figure}

\section{RESULT}

We first study the elementary excitations of the model 
to clarify how the interactions modify the low-energy 
properties of the system.
We calculate the excitation gap for the
charge ($\Delta_n$), spin ($\Delta_s$), and channel
($\Delta_\tau$) degrees of freedom in the parameter space of $J_s$-$J_\tau$.
Figure \ref{fig:ext_gap} shows the excitation gaps obtained
along the line defined by Eq.~(\ref{eq:interaction}).
With decreasing the parameter $J$ of the TCKLM (with increasing $J_s$ and $J_\tau$
of the $t$-$J_s$-$J_\tau$ model),
the spin excitation gap first opens, and then, the charge and channel gaps open.

These successive transitions show the presence of the spin-gap phase.
To further confirm the spin-gap phase, we systematically calculate
the excitation gaps for various $J_s$ and $J_\tau$ and
determine the ground-state phase diagram of the $t$-$J_s$-$J_\tau$ model.
Figure \ref{fig:phase_diagram}
shows the obtained phase diagram consisting of five phases:
metallic phase (no excitation gap),
spin-gap phase (only the spin gap opens),
channel-gap phase (only the channel gap opens),
insulating phase (gap opens for all excitations),
and phase separation.
From the diagram, it is confirmed that the spin-gap phase is 
realized in the TCKLM between the metallic and insulating phases.

As shown in Fig.~\ref{fig:phase_diagram},
the transition lines are symmetric with respect to the line of $J_s=J_\tau$.
This arises from the invariance of the Hamiltonian of Eq.~(\ref{eq:effective-model}) 
and the particle filling $n=1/2$ under the transformation
$(a^\dagger_{i\uparrow}, a^\dagger_{i\downarrow},b^\dagger_{i1}, b^\dagger_{i2}) \rightarrow
(b^\dagger_{i1}, b^\dagger_{i2}, a^\dagger_{i\uparrow}, a^\dagger_{i\downarrow})$
with the exchange of $J_s$ and $J_\tau$,
where the symmetry of the repulsive term $V$ is ensured by the condition of Eq.~(\ref{eq:ed-condition}) 
\footnote{
In the system under the open boundary condition,
the repulsive term modifies
the local chemical potential at both ends of the system.
In this study, we use the SSD and remove such a chemical-potential difference. 
}.
Since this transformation exchanges the role of spin and channel degrees of freedom,
the symmetric phase diagram is obtained.
In the parameter sets we have studied, the direct transition between
the metallic phase and the insulating phase occurs only on the
line of $J_s = J_\tau$.
This implies the existence of a quantum tetracritical point.

\begin{figure}[t]
 \centering
 \includegraphics[scale=1]{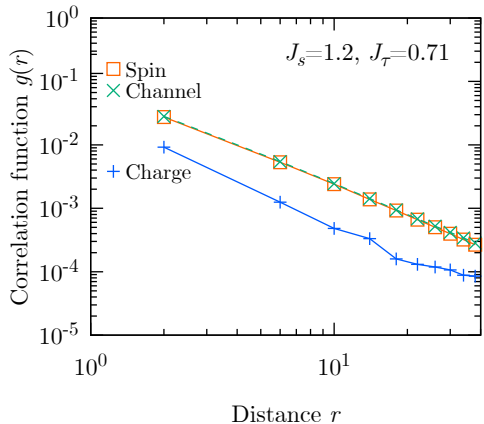}
 \caption{Correlation functions in the metallic phase at $J_s = 1.2$ and $J_\tau= 0.71$ ($J=9.0$).}
 \label{fig:metal_long_corr}
\end{figure}

\subsection{Metallic phase}

\begin{figure}[t]
 \centering
 \includegraphics[scale=1]{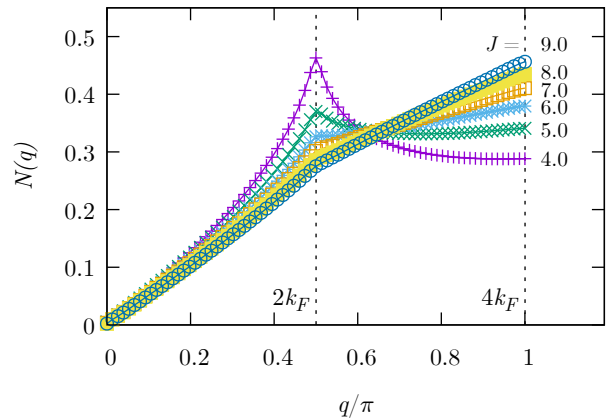}
 \caption{Fourier components of the charge correlation function.
 Central $L=152$ sites of a 192-site system are used to suppress the boundary effects.
 The dominant wavelength changes from $4k_F$ to $2k_F$ with the decrease in $J$ (with the 
 increase in $J_s$ and $J_\tau$).}
 \label{fig:metal_corr_fourier}
\end{figure}

Here we focus on the metallic phase in the region of weak interaction,
where all the excitations are gapless and each spin, charge, and channel degree of freedom behaves as a
Tomonaga-Luttinger liquid (TLL) \cite{cite:haldane_tll, cite:voit_tll}.
As shown in Fig.~\ref{fig:metal_long_corr},
the correlation functions defined by
\begin{align}
 g(r) &= \braket{X_j X_{j+r}} - \braket{X_j}\braket{X_{j+r}},\\
 X_j &= \begin{cases}
  n_j & \text{(for charge)}, \\
  S^z_j & \text{(for spin)}, \\
  \tau^z_j & \text{(for channel)}
 \end{cases}
\end{align}
decay in a power-law fashion $r^{-\alpha}$.
For spin and channel degrees of freedom, the exponent is $\alpha\sim 1.6$ at $J=9$,
which is almost consistent with the prediction of TLL theory $\alpha=1+K_\rho$, where the 
Luttinger parameter $K_\rho$ is determined as $K_\rho \sim 0.5$ from the slope 
of the Fourier components of the charge correlation function $N(q)$
near $q=0$ \cite{cite:tj_1d_pd, cite:tll-lp, cite:tll-hubbard}.
The $J_s$ dependence of $N(q)$ defined by
\begin{align}
 N(q) = \frac{1}{L} \sum_{i,j=1}^L e^{iq(x_i-x_j)}\left(
 \braket{n_i n_j}-\braket{n_i}\braket{n_j}
 \right) \label{eq:fourier-spectra}
\end{align}
also shows that the period of the charge correlation function
clearly changes from $4k_F$ (two site) to $2k_F$ (four site) 
with the increase in $J_s$ and $J_\tau$ as
presented in Fig.~\ref{fig:metal_corr_fourier}.

When $J_s$ exceeds a critical value, the system undergoes the transition to the spin-gap phase. 
At $J_\tau = 0$, the critical value of $J_s$ is close to the bandwidth $4t$.
Figure~\ref{fig:phase_diagram} shows that this critical value becomes smaller with the increases in $J_\tau$, which indicates the interaction acting on the channel degree of freedom stabilizes the spin-gap phase.
We note that the critical value is insensitive to $V$ when $V$ is sufficiently smaller than $4t$.

\subsection{Spin-gap phase}
\begin{figure}[t]
 \centering
 \includegraphics[scale=1]{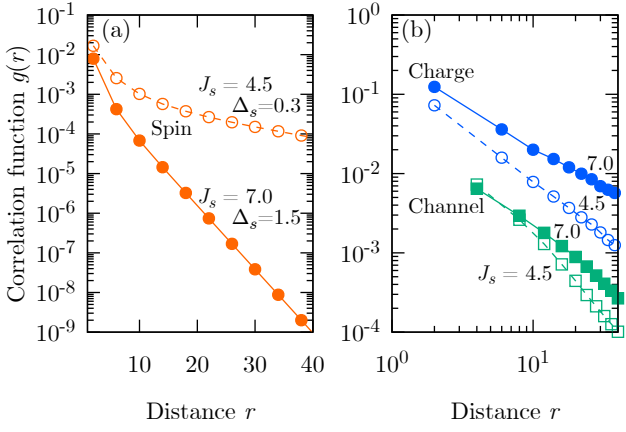}
 \caption{Correlation functions in the spin-gap phase at $J_\tau=0.5$.
 (a) Spin correlation function.
 (b) Charge and channel correlation functions.
 }
 \label{fig:sg_long_corr}
\end{figure}

As discussed above,
the increase in $J_s$ and $J_\tau$ enhances the spin gap,
which makes the slope of the exponential decay of the spin correlation function steeper, as shown in Fig.~\ref{fig:sg_long_corr}(a).
For the charge and channel correlation functions, 
the power-law behavior is confirmed, as seen in Fig.~\ref{fig:sg_long_corr}(b).
The power-law exponent of the charge correlation function
slightly decreases with the increase in the spin gap.

\subsection{Insulating phase}
\begin{figure}[tb]
 \centering
 \includegraphics[scale=1]{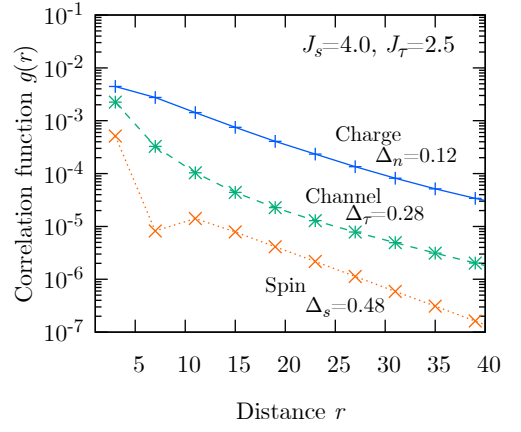}
 \caption{Correlation functions in the insulating phase at
 $J_s=4.0$ and $J_\tau= 2.5$.
}
 \label{fig:insu_long_corr}
\end{figure}

We finally investigate the insulating state.
In the insulating phase, all the excitations 
have a finite energy gap, and the correlation functions decay
exponentially, as shown in Fig. \ref{fig:insu_long_corr}, 
where we find almost the same slope,
although the charge gap is much smaller than the spin gap.
We think this is a result of the alternating product state
wave function of the spin and channel singlets, as shown later.

\begin{figure}[tb]
 \centering
 \includegraphics[scale=1]{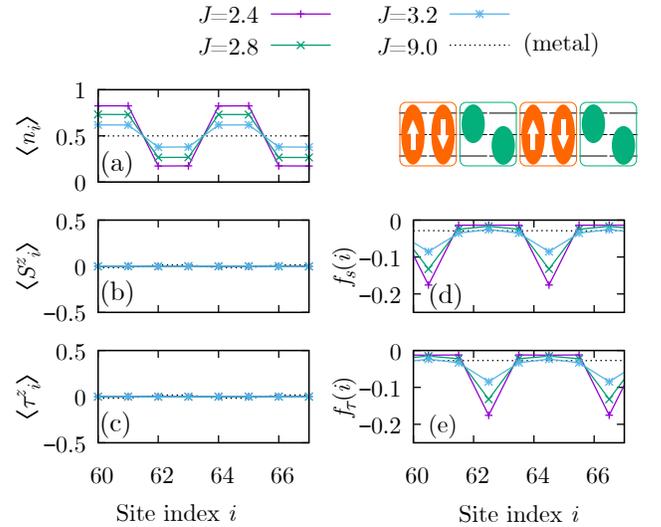}
 \caption{Site-dependent expectation values in
 the insulating phase at $J=2.4$, 2.8, and 3.2.
 (a), (b), and (c) represent local densities.
 (d) and (e) show the nearest-neighbor correlations.
 The dotted line shows those in the
 metallic phase at $J=9.0$ for comparison.
 The top right diagram shows a
 schematic picture of the ground state.
 The rounded rectangles represent singlet pairs.}
 \label{fig:insu_charge_dist_corr}
\end{figure}

To find the symmetry-breaking order of the insulating phase,
we calculate several local expectation values.
Figure \ref{fig:insu_charge_dist_corr} shows the site dependence of the 
local densities and nearest-neighbor correlations, defined as
\begin{align}
 f_s(i) &= \braket{S_{i-1/2}^z S_{i+1/2}^z} - \braket{S_{i-1/2}^z}\braket{S_{i+1/2}^z},\\
 f_\tau(i) &= \braket{\tau_{i-1/2}^z \tau_{i+1/2}^z} - \braket{\tau_{i-1/2}^z}\braket{\tau_{i+1/2}^z},
\end{align}
where $i$ is the center position of the two operators.
We find the charge density $\braket{n_i}$ has 2$k_F$ (four site) oscillation,
whereas the spin and channel densities $\braket{S^z_i}$ and $\braket{\tau^z_i}$ remain zero everywhere.
In addition, the nearest-neighbor correlations strongly correlate
with the charge density oscillation.
These results suggest that the insulating phase is a product state of 
spin and channel singlets, as schematically shown in Fig.~\ref{fig:insu_charge_dist_corr}.
The spin-gap state is then considered as a state in which only the spin degree of freedom forms singlet pairs.

As shown in Fig.~\ref{fig:phase_diagram}, the transition to the insulating state and the opening of the channel gap simultaneously occur in the region of $J_s > J_\tau$.
With the increase in $J_s$ from zero, the critical value of $J_\tau$ which opens the channel gap decreases from almost the bandwidth of $4t$ to $t$ but never goes down to zero, which indicates the cooperation of the spin and channel degrees of freedom is essential for the emergence of the insulating phase.

Here we comment on the effect of the nearest-neighbor repulsion $V$.
This term is added to effectively include the 
higher-order interactions existing in the original TCKLM, which suppress the transition to the phase separation.
As the nearest-neighbor repulsion leads to the metal-insulator transition
in the extended Hubbard model at quarter filling \cite{cite:ext-hubbard1, cite:ext-hubbard2},
this may affect the phase diagram. 
However, the insulating state caused by the nearest-neighbor repulsion $V$
is characterized by the $4k_F$ (2 sites) charge densities, 
which is clearly different from the insulating state found in
the present study, where only $2k_F$ (4 sites) oscillation appears.
We therefore think the repulsion term is not essential in the present analysis.

\section{CONCLUSION}
We have studied the ground states of
the $t$-$J_s$-$J_\tau$ model, which is a minimal model consisting 
of multiple degrees of freedom.
The low-energy excitations of the spin, charge and channel 
degrees of freedom have been calculated by the DMRG method with the SSD, and
it was shown that the phase transition occurs from the metallic state 
to the spin-gap or channel-gap state 
when the exchange interactions exceed almost the bandwidth, 
roughly $\sqrt{J_s^2+J_\tau^2} \sim 4t$.
For the symmetric case of $J_s=J_\tau$, however,
the direct transition to the insulating state takes place.
These results imply that weak channel fluctuations 
stabilize the spin-gap state of the $t$-$J$ model, while
strong channel fluctuations lead to the 
transition to the insulating state which is characterized by the 
alternating product state of the spin and channel singlets.

\section*{ACKNOWLEDGEMENT}
This work was supported by JSPS KAKENHI Grant No.~JP19K03708.

\bibliography{reference}

\begin{thebibliography}{19}%
\makeatletter
\providecommand \@ifxundefined [1]{%
 \@ifx{#1\undefined}
}%
\providecommand \@ifnum [1]{%
 \ifnum #1\expandafter \@firstoftwo
 \else \expandafter \@secondoftwo
 \fi
}%
\providecommand \@ifx [1]{%
 \ifx #1\expandafter \@firstoftwo
 \else \expandafter \@secondoftwo
 \fi
}%
\providecommand \natexlab [1]{#1}%
\providecommand \enquote  [1]{``#1''}%
\providecommand \bibnamefont  [1]{#1}%
\providecommand \bibfnamefont [1]{#1}%
\providecommand \citenamefont [1]{#1}%
\providecommand \href@noop [0]{\@secondoftwo}%
\providecommand \href [0]{\begingroup \@sanitize@url \@href}%
\providecommand \@href[1]{\@@startlink{#1}\@@href}%
\providecommand \@@href[1]{\endgroup#1\@@endlink}%
\providecommand \@sanitize@url [0]{\catcode `\\12\catcode `\$12\catcode
  `\&12\catcode `\#12\catcode `\^12\catcode `\_12\catcode `\%12\relax}%
\providecommand \@@startlink[1]{}%
\providecommand \@@endlink[0]{}%
\providecommand \url  [0]{\begingroup\@sanitize@url \@url }%
\providecommand \@url [1]{\endgroup\@href {#1}{\urlprefix }}%
\providecommand \urlprefix  [0]{URL }%
\providecommand \Eprint [0]{\href }%
\providecommand \doibase [0]{https://doi.org/}%
\providecommand \selectlanguage [0]{\@gobble}%
\providecommand \bibinfo  [0]{\@secondoftwo}%
\providecommand \bibfield  [0]{\@secondoftwo}%
\providecommand \translation [1]{[#1]}%
\providecommand \BibitemOpen [0]{}%
\providecommand \bibitemStop [0]{}%
\providecommand \bibitemNoStop [0]{.\EOS\space}%
\providecommand \EOS [0]{\spacefactor3000\relax}%
\providecommand \BibitemShut  [1]{\csname bibitem#1\endcsname}%
\let\auto@bib@innerbib\@empty
\bibitem [{\citenamefont {Zhang}\ and\ \citenamefont
  {Rice}(1988)}]{cite:zhang_rice}%
  \BibitemOpen
  \bibfield  {author} {\bibinfo {author} {\bibfnamefont {F.~C.}\ \bibnamefont
  {Zhang}}\ and\ \bibinfo {author} {\bibfnamefont {T.~M.}\ \bibnamefont
  {Rice}},\ }\href {https://doi.org/10.1103/PhysRevB.37.3759} {\bibfield
  {journal} {\bibinfo  {journal} {Phys. Rev. B}\ }\textbf {\bibinfo {volume}
  {37}},\ \bibinfo {pages} {3759} (\bibinfo {year} {1988})}\BibitemShut
  {NoStop}%
\bibitem [{\citenamefont {Moreno}\ \emph {et~al.}(2011)\citenamefont {Moreno},
  \citenamefont {Muramatsu},\ and\ \citenamefont {Manmana}}]{cite:tj_1d_pd}%
  \BibitemOpen
  \bibfield  {author} {\bibinfo {author} {\bibfnamefont {A.}~\bibnamefont
  {Moreno}}, \bibinfo {author} {\bibfnamefont {A.}~\bibnamefont {Muramatsu}},\
  and\ \bibinfo {author} {\bibfnamefont {S.~R.}\ \bibnamefont {Manmana}},\
  }\href {https://doi.org/10.1103/PhysRevB.83.205113} {\bibfield  {journal}
  {\bibinfo  {journal} {Phys. Rev. B}\ }\textbf {\bibinfo {volume} {83}},\
  \bibinfo {pages} {205113} (\bibinfo {year} {2011})}\BibitemShut {NoStop}%
\bibitem [{\citenamefont {Troyer}\ \emph {et~al.}(1993)\citenamefont {Troyer},
  \citenamefont {Tsunetsugu}, \citenamefont {Rice}, \citenamefont {Riera},\
  and\ \citenamefont {Dagotto}}]{cite:troyer}%
  \BibitemOpen
  \bibfield  {author} {\bibinfo {author} {\bibfnamefont {M.}~\bibnamefont
  {Troyer}}, \bibinfo {author} {\bibfnamefont {H.}~\bibnamefont {Tsunetsugu}},
  \bibinfo {author} {\bibfnamefont {T.~M.}\ \bibnamefont {Rice}}, \bibinfo
  {author} {\bibfnamefont {J.}~\bibnamefont {Riera}},\ and\ \bibinfo {author}
  {\bibfnamefont {E.}~\bibnamefont {Dagotto}},\ }\href
  {https://doi.org/10.1103/PhysRevB.48.4002} {\bibfield  {journal} {\bibinfo
  {journal} {Phys. Rev. B}\ }\textbf {\bibinfo {volume} {48}},\ \bibinfo
  {pages} {4002} (\bibinfo {year} {1993})}\BibitemShut {NoStop}%
\bibitem [{\citenamefont {Cox}(1987)}]{cite:cox}%
  \BibitemOpen
  \bibfield  {author} {\bibinfo {author} {\bibfnamefont {D.~L.}\ \bibnamefont
  {Cox}},\ }\href {https://doi.org/10.1103/PhysRevLett.59.1240} {\bibfield
  {journal} {\bibinfo  {journal} {Phys. Rev. Lett.}\ }\textbf {\bibinfo
  {volume} {59}},\ \bibinfo {pages} {1240} (\bibinfo {year}
  {1987})}\BibitemShut {NoStop}%
\bibitem [{\citenamefont {Onimaru}\ and\ \citenamefont
  {Kusunose}(2016)}]{cite:onimaru}%
  \BibitemOpen
  \bibfield  {author} {\bibinfo {author} {\bibfnamefont {T.}~\bibnamefont
  {Onimaru}}\ and\ \bibinfo {author} {\bibfnamefont {H.}~\bibnamefont
  {Kusunose}},\ }\href {https://doi.org/10.7566/JPSJ.85.082002} {\bibfield
  {journal} {\bibinfo  {journal} {J. Phys. Soc. Jpn}\ }\textbf {\bibinfo
  {volume} {85}},\ \bibinfo {pages} {082002} (\bibinfo {year}
  {2016})}\BibitemShut {NoStop}%
\bibitem [{\citenamefont {Nozi\`eres}\ and\ \citenamefont
  {Blandin}(1980)}]{cite:nozieres}%
  \BibitemOpen
  \bibfield  {author} {\bibinfo {author} {\bibfnamefont {P.}~\bibnamefont
  {Nozi\`eres}}\ and\ \bibinfo {author} {\bibfnamefont {A.}~\bibnamefont
  {Blandin}},\ }\href {https://doi.org/10.1051/jphys:01980004103019300}
  {\bibfield  {journal} {\bibinfo  {journal} {J. Phys. (Paris)}\ }\textbf
  {\bibinfo {volume} {41}},\ \bibinfo {pages} {193} (\bibinfo {year}
  {1980})}\BibitemShut {NoStop}%
\bibitem [{\citenamefont {Schauerte}\ \emph {et~al.}(2005)\citenamefont
  {Schauerte}, \citenamefont {Cox}, \citenamefont {Noack}, \citenamefont {van
  Dongen},\ and\ \citenamefont {Batista}}]{cite:schauerte}%
  \BibitemOpen
  \bibfield  {author} {\bibinfo {author} {\bibfnamefont {T.}~\bibnamefont
  {Schauerte}}, \bibinfo {author} {\bibfnamefont {D.~L.}\ \bibnamefont {Cox}},
  \bibinfo {author} {\bibfnamefont {R.~M.}\ \bibnamefont {Noack}}, \bibinfo
  {author} {\bibfnamefont {P.~G.~J.}\ \bibnamefont {van Dongen}},\ and\
  \bibinfo {author} {\bibfnamefont {C.~D.}\ \bibnamefont {Batista}},\ }\href
  {https://doi.org/10.1103/PhysRevLett.94.147201} {\bibfield  {journal}
  {\bibinfo  {journal} {Phys. Rev. Lett.}\ }\textbf {\bibinfo {volume} {94}},\
  \bibinfo {pages} {147201} (\bibinfo {year} {2005})}\BibitemShut {NoStop}%
\bibitem [{\citenamefont {White}(1992)}]{cite:dmrg_white1}%
  \BibitemOpen
  \bibfield  {author} {\bibinfo {author} {\bibfnamefont {S.~R.}\ \bibnamefont
  {White}},\ }\href {https://doi.org/10.1103/PhysRevLett.69.2863} {\bibfield
  {journal} {\bibinfo  {journal} {Phys. Rev. Lett.}\ }\textbf {\bibinfo
  {volume} {69}},\ \bibinfo {pages} {2863} (\bibinfo {year}
  {1992})}\BibitemShut {NoStop}%
\bibitem [{\citenamefont {White}(1993)}]{cite:dmrg_white2}%
  \BibitemOpen
  \bibfield  {author} {\bibinfo {author} {\bibfnamefont {S.~R.}\ \bibnamefont
  {White}},\ }\href {https://doi.org/10.1103/PhysRevB.48.10345} {\bibfield
  {journal} {\bibinfo  {journal} {Phys. Rev. B}\ }\textbf {\bibinfo {volume}
  {48}},\ \bibinfo {pages} {10345} (\bibinfo {year} {1993})}\BibitemShut
  {NoStop}%
\bibitem [{\citenamefont {Gendiar}\ \emph {et~al.}(2009)\citenamefont
  {Gendiar}, \citenamefont {Krcmar},\ and\ \citenamefont
  {Nishino}}]{cite:ssd_gendiar1}%
  \BibitemOpen
  \bibfield  {author} {\bibinfo {author} {\bibfnamefont {A.}~\bibnamefont
  {Gendiar}}, \bibinfo {author} {\bibfnamefont {R.}~\bibnamefont {Krcmar}},\
  and\ \bibinfo {author} {\bibfnamefont {T.}~\bibnamefont {Nishino}},\ }\href
  {https://doi.org/10.1143/PTP.122.953} {\bibfield  {journal} {\bibinfo
  {journal} {Prog. Theor. Phys.}\ }\textbf {\bibinfo {volume} {122}},\ \bibinfo
  {pages} {953} (\bibinfo {year} {2009})}\BibitemShut {NoStop}%
\bibitem [{\citenamefont {Gendiar}\ \emph {et~al.}(2010)\citenamefont
  {Gendiar}, \citenamefont {Krcmar},\ and\ \citenamefont
  {Nishino}}]{cite:ssd_gendiar2}%
  \BibitemOpen
  \bibfield  {author} {\bibinfo {author} {\bibfnamefont {A.}~\bibnamefont
  {Gendiar}}, \bibinfo {author} {\bibfnamefont {R.}~\bibnamefont {Krcmar}},\
  and\ \bibinfo {author} {\bibfnamefont {T.}~\bibnamefont {Nishino}},\ }\href
  {https://doi.org/10.1143/PTP.123.393} {\bibfield  {journal} {\bibinfo
  {journal} {Prog. Theor. Phys.}\ }\textbf {\bibinfo {volume} {123}},\ \bibinfo
  {pages} {393} (\bibinfo {year} {2010})}\BibitemShut {NoStop}%
\bibitem [{\citenamefont {Hotta}\ and\ \citenamefont
  {Shibata}(2012)}]{cite:ssd_hotta}%
  \BibitemOpen
  \bibfield  {author} {\bibinfo {author} {\bibfnamefont {C.}~\bibnamefont
  {Hotta}}\ and\ \bibinfo {author} {\bibfnamefont {N.}~\bibnamefont
  {Shibata}},\ }\href {https://doi.org/10.1103/PhysRevB.86.041108} {\bibfield
  {journal} {\bibinfo  {journal} {Phys. Rev. B}\ }\textbf {\bibinfo {volume}
  {86}},\ \bibinfo {pages} {041108(R)} (\bibinfo {year} {2012})}\BibitemShut
  {NoStop}%
\bibitem [{Note1()}]{Note1}%
  \BibitemOpen
  \bibinfo {note} {In the system under the open boundary condition, the
  repulsive term modifies the local chemical potential at both ends of the
  system. In this study, we use the SSD and remove such a chemical-potential
  difference.}\BibitemShut {Stop}%
\bibitem [{\citenamefont {Haldane}(1981)}]{cite:haldane_tll}%
  \BibitemOpen
  \bibfield  {author} {\bibinfo {author} {\bibfnamefont {F.~D.~M.}\
  \bibnamefont {Haldane}},\ }\href
  {https://doi.org/10.1088/0022-3719/14/19/010} {\bibfield  {journal} {\bibinfo
   {journal} {J. Phys. C}\ }\textbf {\bibinfo {volume} {14}},\ \bibinfo {pages}
  {2585} (\bibinfo {year} {1981})}\BibitemShut {NoStop}%
\bibitem [{\citenamefont {Voit}(1995)}]{cite:voit_tll}%
  \BibitemOpen
  \bibfield  {author} {\bibinfo {author} {\bibfnamefont {J.}~\bibnamefont
  {Voit}},\ }\href {https://doi.org/10.1088/0034-4885/58/9/002} {\bibfield
  {journal} {\bibinfo  {journal} {Rep. Prog. Phys}\ }\textbf {\bibinfo {volume}
  {58}},\ \bibinfo {pages} {977} (\bibinfo {year} {1995})}\BibitemShut
  {NoStop}%
\bibitem [{\citenamefont {Clay}\ \emph {et~al.}(1999)\citenamefont {Clay},
  \citenamefont {Sandvik},\ and\ \citenamefont {Campbell}}]{cite:tll-lp}%
  \BibitemOpen
  \bibfield  {author} {\bibinfo {author} {\bibfnamefont {R.~T.}\ \bibnamefont
  {Clay}}, \bibinfo {author} {\bibfnamefont {A.~W.}\ \bibnamefont {Sandvik}},\
  and\ \bibinfo {author} {\bibfnamefont {D.~K.}\ \bibnamefont {Campbell}},\
  }\href {https://doi.org/10.1103/PhysRevB.59.4665} {\bibfield  {journal}
  {\bibinfo  {journal} {Phys. Rev. B}\ }\textbf {\bibinfo {volume} {59}},\
  \bibinfo {pages} {4665} (\bibinfo {year} {1999})}\BibitemShut {NoStop}%
\bibitem [{\citenamefont {Schulz}(1990)}]{cite:tll-hubbard}%
  \BibitemOpen
  \bibfield  {author} {\bibinfo {author} {\bibfnamefont {H.~J.}\ \bibnamefont
  {Schulz}},\ }\href {https://doi.org/10.1103/PhysRevLett.64.2831} {\bibfield
  {journal} {\bibinfo  {journal} {Phys. Rev. Lett.}\ }\textbf {\bibinfo
  {volume} {64}},\ \bibinfo {pages} {2831} (\bibinfo {year}
  {1990})}\BibitemShut {NoStop}%
\bibitem [{\citenamefont {Mila}\ and\ \citenamefont
  {Zotos}(1993)}]{cite:ext-hubbard1}%
  \BibitemOpen
  \bibfield  {author} {\bibinfo {author} {\bibfnamefont {F.}~\bibnamefont
  {Mila}}\ and\ \bibinfo {author} {\bibfnamefont {X.}~\bibnamefont {Zotos}},\
  }\href {https://doi.org/10.1209/0295-5075/24/2/010} {\bibfield  {journal}
  {\bibinfo  {journal} {Europhys. Lett.}\ }\textbf {\bibinfo {volume} {24}},\
  \bibinfo {pages} {133} (\bibinfo {year} {1993})}\BibitemShut {NoStop}%
\bibitem [{\citenamefont {Shirakawa}\ and\ \citenamefont
  {Jeckelmann}(2009)}]{cite:ext-hubbard2}%
  \BibitemOpen
  \bibfield  {author} {\bibinfo {author} {\bibfnamefont {T.}~\bibnamefont
  {Shirakawa}}\ and\ \bibinfo {author} {\bibfnamefont {E.}~\bibnamefont
  {Jeckelmann}},\ }\href {https://doi.org/10.1103/PhysRevB.79.195121}
  {\bibfield  {journal} {\bibinfo  {journal} {Phys. Rev. B}\ }\textbf {\bibinfo
  {volume} {79}},\ \bibinfo {pages} {195121} (\bibinfo {year}
  {2009})}\BibitemShut {NoStop}%
\end{thebibliography}%
\end{document}